# Optical Pulling and Pushing Forces via Bloch Surface Waves

Natalia Kostina[1]*, Mihail Petrov[1], Vjaceslavs Bobrovs[2], Alexander S. Shalin[1,3,4]

[1] *ITMO University, Saint-Petersburg, Russia, Kronverksky prospekt 49, 197101*

[2] *Institute of Telecommunications, Riga, Latvia, Azenes street 12, 1048*

[3] *Institute of Nanotechnology of Microelectronics of the Russian Academy of Sciences (INME RAS), Moscow, Russia, Nagatinskaya street 16A, bldg. 11, 119991*

[4] *Kotel'nikov Institute of Radio Engineering and Electronics of Russian Academy of Sciences (Ulyanovsk branch), Ulyanovsk, Russia, Goncharova street 48, 432000*

*natalia.kostina@metalab.ifmo.ru

**Abstract:** Versatile manipulation of nano- and microobjects underlies the optomechanics and a variety of its applications in biology, medicine, and lab-on-a-chip platforms. For flexible tailoring optical forces, as well as for extraordinary optomechanical effects, additional degrees of freedom should be introduced into the system. Here, we demonstrate that photonic crystals provide a flexible platform for nanoparticles optical manipulation due to both Bloch surface waves (BSWs) and the complex character of the reflection coefficient paving a way for complex optomechanical interactions control. We demonstrate that appearance of enhanced pulling and pushing transversal optical forces acting on a single bead placed above a one-dimensional photonic crystal due to directional excitation of Bloch surface wave at the photonic crystal interface. Our theoretical results, which are supported with numerical simulations, demonstrate angle or wavelength assisted switching between BSW-induced optical pulling and pushing forces. Easy-to-fabricate for any desired spectral range photonic crystals are shown to be prospective for precise optical sorting of nanoparticles, especially for core-shell nanoparticles, which are difficult to sort with conventional optomechanical methods. Our approach opens opportunities for novel optical manipulation schemes and platforms and enhanced light-matter interaction in optical trapping setups.

## I Introduction

The seminal papers on optical forces by Arthur Ashkin [1,2] gave rise to an ever-increasing research in optomechanics and optical manipulation as well as inspired a variety of applications in biology[3–5], lab-on-a-chip platforms[6–10], micro- and nanooptomechanical devices[11–17], and others. One of the mostly demanded optomechanical applications is the micro- and nanoparticles' trapping and transport along an arbitrary predefined trajectory [18–24]. Conventionally, it can be achieved by direct movement of an optical tweezer, a focused laser beam, with a trapped particle, however, that requires using special optical equipment such as spatial light modulator and precise positioning systems. Alternatively, one can access the novel tools offered by modern nanophotonics which provide new degrees of freedom in optical manipulation via specially designed auxiliary nanostructures and engineered scattering channels. For example, optical pulling forces ('tractor beams') can be obtained for particles with a complex multipolar response irradiated with so-called non-diffractive beams (e.g., Bessel beams) [6,25–27], or on air-water interfaces[28] via a momentum difference of the incident and scattered light.

Since 2015-2016 it has been suggested that the pulling optical forces can be also driven by the substrate modes and the studies showed the tailored particles attraction by directional excitation of guided modes of a dielectric waveguide[29] and optical pulling of a dipole particle via surface plasmon-polaritons[30,31]. At the same time, multilayered structures can support different types of modes depending on the actual structure and materials which opens more possibilities for engineering their optical properties. The unique optical forces, resulting from the use of auxiliary substrates, provided plasmonic-induced enhanced trapping and antitrapping[32,33], extra-strong subwavelength plasmonic interactions of several particles[34–36], and chromatically tunable binding featured by hyperbolic modes of a multilayered metal-dielectric slab[37]. The optical pulling force driven by volumetric modes of a hyperbolic metamaterial was predicted[38], while strong attractive and repulsive optical forces were obtained on a particle near a sandwich metal-dielectric-metal substrate due to flat-band plasmon mode excitation[39]. However, the utilization of metallic substrates meets particular limitations such as strong optical losses resulting in thermal heating of samples. To avoid that, one-dimensional photonic crystal (1D PC) have been suggested for the enhancing particle's propulsion[40–42]. This unusual application of 1D PC became possible due to low-loss long-range surface Bloch waves[43] and rather simple fabrication procedure of PCs. Moreover, photonic crystals support high-k modes in an arbitrary optical range and are not restricted by surface plasmon resonance conditions. Currently, BSWs of such crystals are widely used in detection devices[44–48] and high-resolution imaging [46,49–53].

Here, inspired by the recent developments in photonic crystals-enhanced optomechanics[54–56], we focus on the pulling optical forces, forces switching, and optical sorting by tailored directional Bloch surface waves (BSWs) [57,58]. We analyze the self-consistent problem of nanoparticles interaction with 1D PC, which results in a rather diverse behavior of forces. For example, BSW direction strongly depends on the angle and wavelength of incident light, and on the particles' parameters, which drive switching between intensive optical pulling and pushing forces. We consider in detail the modal structure (surface and volumetric modes) of the system and optical forces, demonstrating the conditions for the consequent attractive and repulsive behavior.

## II Optical force calculation

Let us consider the scheme consisting of a single dielectric particle placed on top of a one-dimensional photonic crystal (1D PC), as shown in Figure 1. The system is illuminated by a p-polarized wave at the angle of incidence $\theta$.

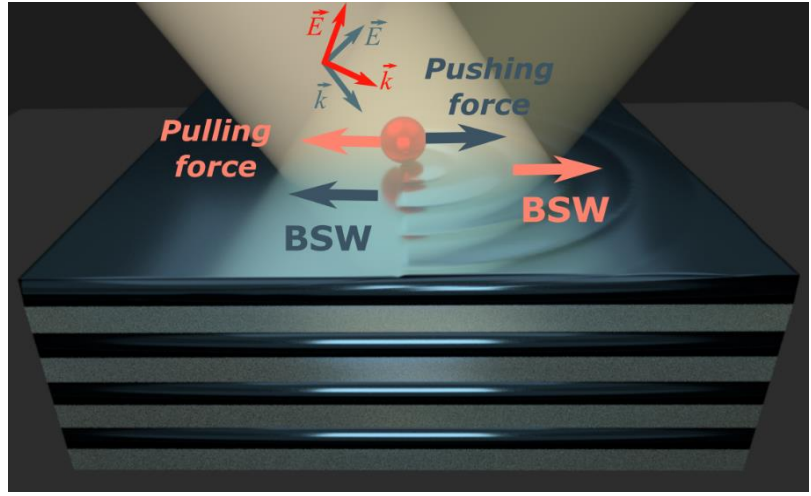

Figure 1 – A dielectric bead is placed on top of a one-dimensional photonic crystal and illuminated by an inclined p-polarized plane wave. Directional excitation of Bloch surface wave (in either forward or backward directions) leads to the recoil force opposite to the BSW propagation.

The optical force acting on a small particle can be written in the following general form [59]:

$$\boldsymbol{F} = \frac{1}{2}\operatorname{Re}\left[\sum_i p_i^* \nabla E_i^{loc}\right], \quad i = x, y, z \qquad (1)$$

where $\boldsymbol{p}^*$ is complex conjugate dipole moment of the particle, $\boldsymbol{E}^{loc}$ is a local field in the dipole position taking into account the field from the incident wave, and field rescattered by the particle. Considering the correction for the particle-substrate interaction[30] one can obtain the following expressions:

$$\boldsymbol{E}^{loc} = \boldsymbol{E}^0 + \boldsymbol{E}^{PS}$$
$$\boldsymbol{p} = \alpha_0\left[\ddot{\boldsymbol{I}} - \alpha_0 \frac{k^2}{\varepsilon_0}\ddot{\boldsymbol{G}}(\vec{r}_0,\vec{r}_0)\right]^{-1} \boldsymbol{E}^0 \qquad (2)$$

$\boldsymbol{E}^0$ is the field from incident and reflected plane wave, $\boldsymbol{E}^{PS}$ is the field induced by particle-substrate interaction, $\alpha_0$ is a dipole polarizability of the particle with the retardation correction, $\ddot{\boldsymbol{I}}$ is a unitary dyad, $k$ is a wavenumber in a free space, $\varepsilon_0$ is the permittivity of vacuum, $\ddot{\boldsymbol{G}}(\vec{r}_0,\vec{r}_0)$ is a dyadic Green's function describing self-induced contribution to the particle's polarizability and $\boldsymbol{E}^{PS}$.

Following [30], we split the lateral optical force into two parts

$$F_x = \frac{1}{2} k \sin(\theta) \operatorname{Im} \left[ \alpha_0 \left[ 1 - \alpha_0 \frac{k^2}{\varepsilon_0} G_{xx} \right]^{-1} |E_x^0|^2 + \alpha_0 \left[ 1 - \alpha_0 \frac{k^2}{\varepsilon_0} G_{zz} \right]^{-1} |E_z^0|^2 \right] - \frac{k^2}{\varepsilon_0} \operatorname{Im}\left[ p_x^* p_z \right] \operatorname{Im} \partial_x G_{xz}, \quad (3)$$

where $\theta$ is the angle of the plane wave incidence. The first term is the pressure force corresponding to the direct action of the incident wave on the polarized dipole, and it is always positive. The second term describes the impact of the particle's self-action through the substrate and takes into account the near fields generated in the vicinity of the substrate surface. Thus, it contains the substrate response and provides all the peculiarities in optomechanical action related to PC mode structure. As seen, the two multipliers determine the second component value and sign: (i) the imaginary part of the Green's function derivative, which is positive (see Supporting Information 1) and defines the magnitude of the force; (ii) the imaginary part of the dipole moments product, which is dependent on the angle of incidence

$$\operatorname{Im}\left[ p_x^* p_z \right] \sim |\alpha_0|^2 |E^0|^2 \sin(2\theta) \operatorname{Im}\left[ r^p(\theta) e^{2ik\cos(\theta)z} \right] \quad (4)$$

and its sign is determined by the reflection coefficient. If we consider small distances between the nanoparticle and the surface, $\operatorname{Im}\left[ r^p(\theta) \right]$ defines the character of forces only [30,60]. For $\operatorname{Im}\left[ r^p(\theta) \right] > 0$ the contribution to the optical force is negative, and the optical pulling effect becomes possible, for $\operatorname{Im}\left[ r^p(\theta) \right] < 0$ the substrate gives a positive component of the force and, consequently, an additional enhancement of optical pushing. In other words, the sign of the imaginary part of reflection coefficient for the incident plane wave defines the sign of phase difference between $p_x$, $p_z$, providing positive or negative dipole moment rotation and corresponding forward or backward BSW propagation (Figure 2) due to selective excitation of surface wave with circulating dipole moment[61–63]. The latter condition also holds for metallic substrates in a narrow region near the surface plasmon resonance, however, it gives insignificant contribution to pure radiation pressure. In the case of a photonic crystal, the reflection coefficient can have more complicated behavior due to Fabry-Perot peaks (Figure 3), which in turn could lead to the pushing-pulling switching.

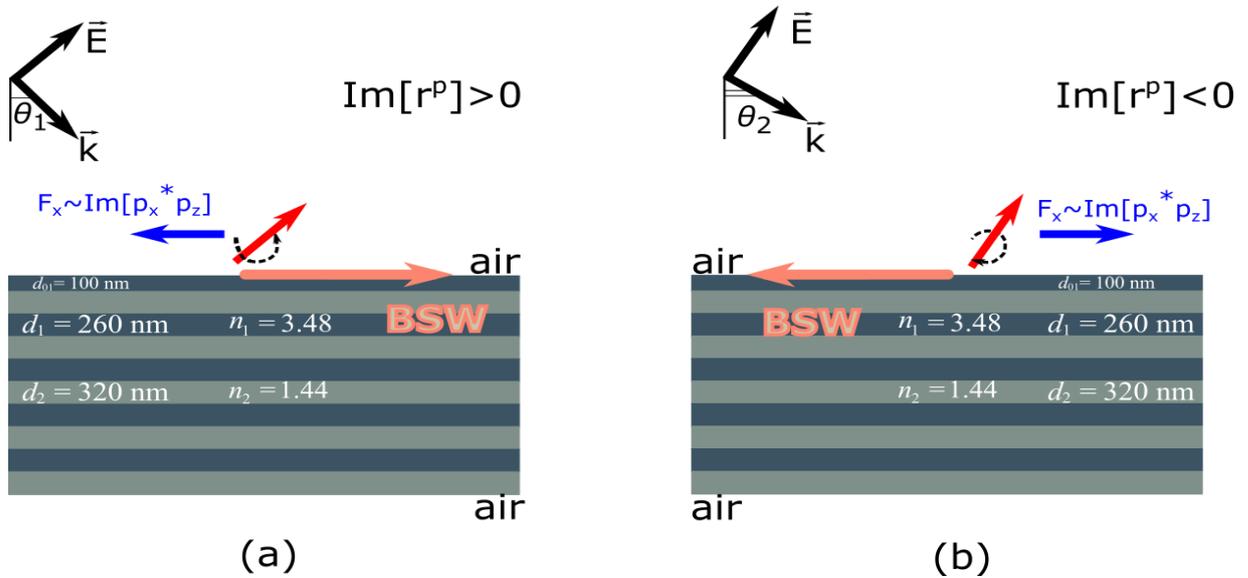

Figure 2 – The principal scheme of the problem. (a) $\operatorname{Im}\left[ r^p(\theta) \right] > 0$ leads to BSW propagation in positive x-direction and negative contribution of the second term in Eq.(3). provides an optical

pulling force; (b) $\text{Im}\left[r^p(\theta)\right] < 0$, and BSW propagates in negative x-direction and the recoil force in this geometry adds to total pushing.

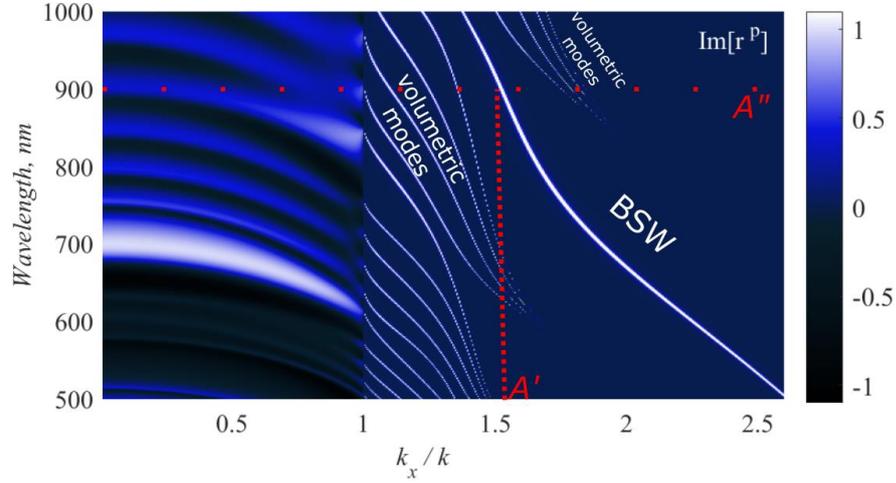

Figure 3– The imaginary part of the reflection coefficient from photonic crystal slab of 5 periods, with refractive index of layers $n_1 = 3.48$, $n_2 = 1.44$, and thicknesses $d_1 = 260\,\text{nm}$, $d_2 = 320\,\text{nm}$. The topmost layer has thickness $d_{01} = 100\,\text{nm}$. In the region of upper half-space propagating modes $k_x \in [0, k]$ negative and positive values alternation is wavelength and angle dependent. In the region of evanescent modes $k_x \in (k, \infty)$, the wide bright peak corresponding to BSW takes place, and several ensembles of narrow peaks show volumetric modes inside the PC. Line A' denotes the mathematical limit of the integration for distinguishing surface and volumetric modes contributions, while A'' shows the wavelengths for such a simplification.

From this point of view, directional excitation of BSW can be achieved when the phase difference between $p_x$ and $p_z$ (or $E_x^0$ and $E_z^0$) reaches $\pm \pi/2$ and their amplitudes become equal providing induced circular polarization of the dipole particle. It, in turn, leads to the asymmetric scattering from the dipole. $\pi/2$ phase difference results in BSW excitation only along positive x-direction (see Figure 2 (a)) and recoiling optical pulling force (if overcomes radiation pressure), in contrast $-\pi/2$ phase results in BSW excitation in negative direction, and therefore recoiling force enhances radiation pressure (see Figure 2 (b)).

BSW contribution to the optical force (independently on its direction) can be easily estimated via the Green's function formalism as its components can be expressed through the integrals over transversal wavevector component $k_x$ over the interval $[0, \infty)$ (See Supporting Information 1 for the exact expressions). Integration over $k_x \in [0, k]$ corresponds to free-space (upper half-space) propagating modes and does not include surface/volumetric modes. On the other hand, we can estimate the contributions of the substrate modes by extracting modes from the interval $k_x \in (k, \infty)$.

## III Results and discussion

Let us consider 1D PC with $N=5$ alternating layers of $n_1=3.48$, $d_1=260$ nm and $n_2=1.44$, $d_2=320$ nm[53]. The topmost layer thickness is $d_{01}=100$ nm to support the minimal amount of substrate modes for simplicity. Figure 3 shows the reflection coefficient of the PC slab as function of wavelength and lateral k-vector showing the structure of the modes. From Figure 3 one can see that up to wavelength of 900 nm BSW wavevector is larger than the one for volumetric modes. We choose limit $k_x=1.51k$ to simply distinguish volumetric and surface modes of the substrate in this spectral range. The red line in the figure corresponds to $k_x=A'=1.51k$, so the integration up to the red line includes free-space and volumetric modes of the substrate, while $k_x>1.51k$ drives the BSW contribution. The residual contribution of volumetric modes beyond the red line limit is insignificant. Though, for longer wavelengths the contributions of bulk and surface waves separation is more cumbersome (see Supporting Information 2).

Figure 4 (a) shows the spectral dependence of the optical force calculated for the angle of incidence $\theta=15°$. The particle with dielectric permittivity $\varepsilon_p=3$ and radius $R=15$ nm is placed above 1D PC. It is clearly seen that the free-space modes and volumetric modes of PC give insignificant optical pulling or pushing forces (smaller than the radiation pressure force from the plane wave acting on the same particle in vacuum $F_0=\frac{1}{2}k\,\mathrm{Im}\,\alpha_0\left|E^{incident}\right|^2$ ). Moreover, the negative optical force (pulling force) is fully governed by the surface modes of the substrate. As mentioned above, the sign of the force is defined by the sign of the phase difference between $p_x$ and $p_z$ or the sign of $\mathrm{Im}\left[r^p(\theta)\right]$. The positive reflection (blue line in Figure 4(b)) corresponds to the optical pulling force (denoted by red regions) and almost $\pi/2$ phase difference. The peaks of the total force and the contributions of the free-space and volumetric modes coincide with the peaks of the imaginary part of the reflection coefficient and $\arg(p_x^* p_z)$ maxima. We note, that there is a slight difference in highlighted regions, when the second summand in Eq.(3) does not completely suppress the first one. The optical pulling/pushing force switching is due to changing the reflection coefficient sign and can be explained as the influence of different propagation directions of BSW at the recoil force sign. Optical force near PC with non-specific topmost layer is shown in Supporting Information 3.

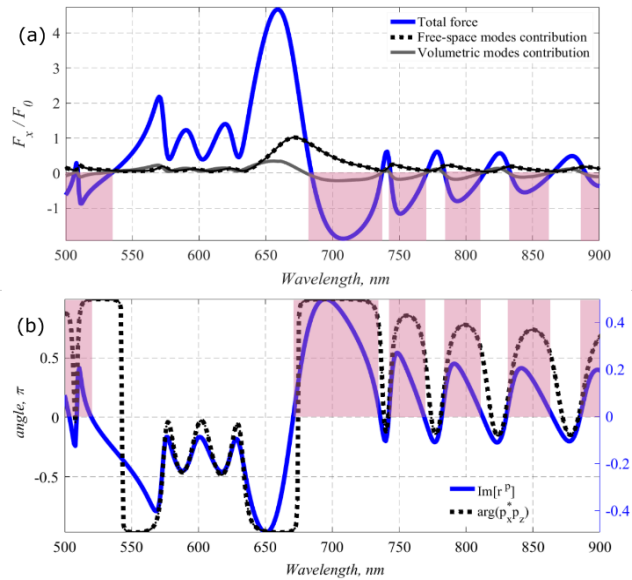

Figure 4 – (a) Optical force spectra: the blue line denotes the total force, the black line denotes free-space modes contribution (Green's function integration upper limit is $k_x=k$ ), the grey line corresponds to the volumetric modes of the PC contribution (Green's function integration upper

limit is $k_x = 1.51k$), i.e. without BSW. The red regions highlight negative total optical force (pulling). All forces are normalized to the radiation pressure force in vacuum at corresponding wavelength $F_0 = \frac{1}{2} k \, \text{Im}\, \alpha_0 \left| E^{incident} \right|^2$. (b) Phase difference between $p_x$ and $p_z$ (black), and the imaginary part of reflection coefficient (blue) for angle of incidence $\theta = 15°$. The red regions highlight positive imaginary part of the reflection coefficient.

Thus, we show that optical force character has a strong dependence on the angle of incidence $\theta$ provided by the reflection coefficient $r^p(\theta)$ and properties of PC slab. It is possible to tune not only the force magnitude, but also the sign of the force by adjusting the angle and wavelength. The colormap in the Figure 5 shows the possibility of switching between optical pulling and pushing behaviors by varying wavelength or angle of incidence and can be tailored by proper designing of PC structure. It is also possible to enhance the pushing force more than ten times in contrast to the surface plasmon polariton based scenario. In Figure 6, we plot optical force spectra for $\theta = 36°, 75°$, showing that the phase difference of the dipole moment components almost reaches $\pi/2$ even for big angles of incidence. However, the region of optical pulling force is narrow for 36o (660-680 nm) (Figure 6(a)) and almost vanishes for 75o nm. It can be explained with a bigger contribution of the first summand in Eq.(3) dependent on $\sin(\theta)$. It is seen from Figure 6 (b) that the compensation of the two summands and, thus, zero transversal optical force can be achieved at around 800 nm.

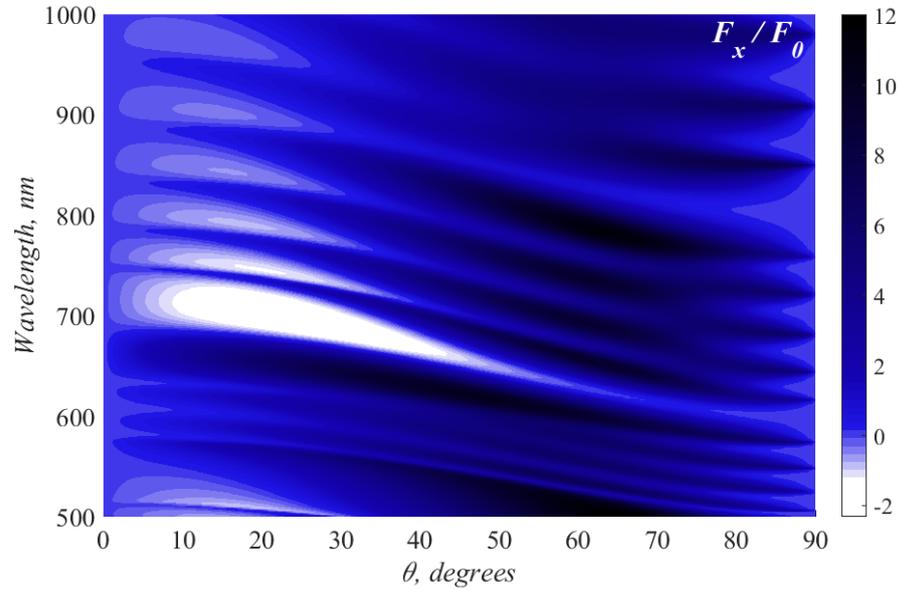

Figure 5 – Map of the optical force dependence on the wavelength and the angle of incidence.

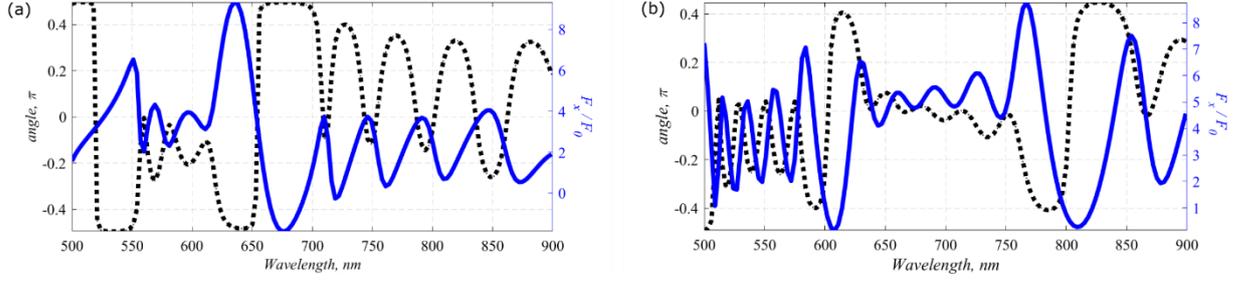

Figure 6 – Optical force (blue) and dipole moment components phase difference (black) for (a) $\theta = 36°$ and (b) $\theta = 75°$.

We compare our analytical results with modelling via Lumerical FDTD Solutions in Figure 7. In the simulation, the small gap should be introduced between the substrate due to mesh requirements. These two curves correspond well, which proves the correctness and completeness of our model.

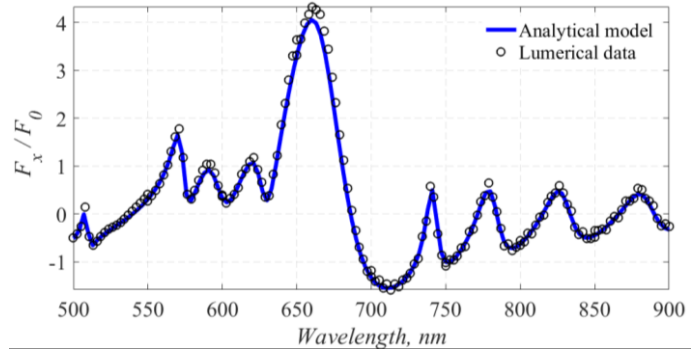

Figure 7 – Comparison of the analytical model (blue line) with the FDTD simulation (black circles). The particle of radius $R = 16$ nm and permittivity $\varepsilon_p = 3$ is placed with center at $z = 20$ nm (4 nm gap between the particle and the substrate).

## IV Sorting applications

We now turn our attention to possible applications of the BSW-assisted force and one of the most prospective on relates to sorting of resonant particles [6,64]. The strong spectral dependence of the pushing/pulling regimes of BSW assisted forces provides the opportunity to separate the particles with the optical resonances falling in particular spectral range disregard of the origin of the resonances. Here, we consider the simplest model of a core-shell plasmonic particles with external radius $R = 40$ nm and varied core radius $R_{core}$ placed on top of the PC. Polarizability of the core-shell particle can be written as [65]:

$$\alpha_0^{core-shell} = R^3 \frac{(\varepsilon_{core} + 2\varepsilon_{shell})(\varepsilon_{shell} - \varepsilon_m) + f(\varepsilon_{core} - \varepsilon_{shell})(\varepsilon_m + 2\varepsilon_{shell})}{(\varepsilon_{core} + 2\varepsilon_{shell})(\varepsilon_{shell} + 2\varepsilon_m) + 2f(\varepsilon_{core} - \varepsilon_{shell})(\varepsilon_{shell} - \varepsilon_m)}, \quad f = \left(\frac{R_{core}}{R}\right)^3 \quad (5)$$

providing the optical resonance, which spectral position strongly depends on the shell thickness. Effective polarizability with respect to particle-substrate interaction could be written as:

$$\hat{\alpha} = \alpha_0 \left[1 - \alpha_0 \frac{k^2}{\varepsilon_0} \hat{G}\right]^{-1}. \quad (6)$$

Figure 8 inset shows the spectra of nanoparticle polarizabilities xx-component for varied core radius with permittivity $\varepsilon_{core} = 2.3$, and the silver shell with permittivity $\varepsilon_{shell}$ given in Ref[66]. Thus, the overall radius of the particle $R_{shell}$ is kept constant, while the radii of the core is varied in the range $R_{core} = 35.5 - 38$ nm.

In Figure 8 the total optical force is shown for nanoparticles of different core radius. With the core radius increasing (reduction of the shell thickness) we tune the resonance position (or switching between negative and positive real part of polarizability) to match PC-induced peaks at longer wavelengths. Thus, $\text{Re}(\alpha_0^{core-shell}) < 0$ changes the optical force sign by adding $\pi$-phase shift to the $\arg(p_x^* p_z)$, and as a result, different shell sizes demonstrate opposite behaviors. It is seen, that the sign change of optical force takes place at 725 nm ($R_{core} = 35.5$ nm optical pulling, $R_{core} = 36$ nm optical pushing), at 750 nm ($R_{core} = 35.5, 36$ nm optical pulling, $R_{core} = 36.5$ nm optical pushing), at around 800 nm ($R_{core} = 36, 36.5$ nm optical pulling, $R_{core} = 37$ nm optical pushing) and at around 850 and 900 nm ($R_{core} = 35.5, 36, 36.5, 37$ nm optical pulling, $R_{core} = 37.5$ nm optical pushing). The most representative examples of optical force sign switching are highlighted by the red dashed rectangles, while more details are available in Supporting Information 4. Thus, by tuning the laser frequency, one can change the pulling/pushing regime for particles of a certain type. Such a scheme opens a way to precise sorting of core-shell particles with sharp spectral change of the optical sorting regions providing good tolerance to the wavelength. To summarize, we demonstrate the possibility to precisely sort resonant nanoparticles via BSWs at different wavelengths, providing the opportunity to discern nanoparticles with relatively small shifts between the resonances. Noteworthy, that this kind of particles are nearly impossible to sort with a conventional optical tweezer.

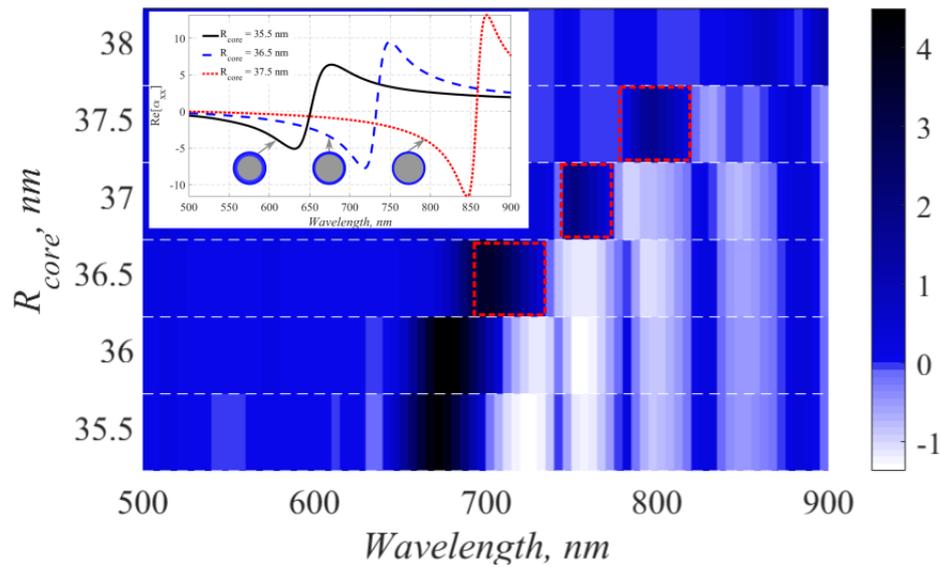

Figure 8 – Optical force acting on a resonant nanoparticle near the PC. White color depicts optical pulling force, black color depicts optical pushing. Red rectangles show regions of optical pulling-to-pushing force alteration. The inset shows real part of the particle polarizability component containing substrate response for $R_{core} = 35.5$ nm (black solid line), $R_{core} = 36.5$ nm (blue dashed line), $R_{core} = 37.5$ nm (red dotted line). Schematics of the particles with different core-to shell ratios are shown with circles.

**Conclusion**

In this paper we propose a novel approach to manipulating nanoobject based on rescattering of Bloch surface wave near one-dimensional photonic crystal. We, for the first time, demonstrate enhanced optical pulling/pushing effect provided by the recoil force appearance due to directional excitation of BSWs. Moreover, we show the possibility for tailoring a sign and magnitude of the optical force by both, incident wavelength and angle of incidence, which, in turn, is defined by the PC parameters and could be easily designed for a desired spectral range. It paves a way to flexible and non-lossy optical manipulation of nanoscale objects and can be used for many applications including nanotechnology, nanochemistry and biology. In particular, we propose sorting mechanism for separating the particles with respect to their optical resonance position such as, for example, core-shell nanoparticles. We show that the particles with different core-shell ratios can be sorted with high precision, which is provided by the optical response of PC structure and cannot be achieved by other conventional approaches, and even in SPPs-enhanced schemes. Thus, the proposed scenario is prospective for optomechanics in wide scope of applications, since it enables versatile control not only over magnitude of the optical force, but also over its sign and can be used to flexible transport of nanoparticles over large distances, that is highly demanded, for instance in lab-on-a-chip and biological applications.

**Acknowledgements**

This work was supported by a grant from the Theoretical Physics and Mathematics Advancement Foundation "BASIS" № 20-1-5-115-1. This work was supported by the Russian Science Foundation (project № 20-72-10141).

# SUPPORTING INFORMATION

## 1. Green's function

The Green's function at the position of a single particle placed above the substrate has the following form [1]:

$$\ddot{G}(r_0, r_0) = \frac{i}{8\pi k^2} \int_0^\infty \frac{k_x}{k_z} \begin{bmatrix} k^2 r^s - k_z^2 r^p & 0 & 0 \\ 0 & k^2 r^s - k_z^2 r^p & 0 \\ 0 & 0 & 2k_x^2 r^p \end{bmatrix} e^{2ik_z z_0} dk_x \quad (7)$$

where $r_0 = (x_0, y_0, z_0)$ denotes the particle's position, $k$ wavenumber of incident light in upper half-space, $k_x, k_z$ transversal and longitudinal components of the wavevector, $r^s, r^p$ reflection coefficient for s- and p-polarized light from the substrate, $z_0$ distance between the substrate and center of the particle. This expression is integrated over propagating in upper half-space modes $k_x \in [0, k]$ and evanescent modes with larger wavenumbers $k_x \in (k, \infty)$.

x-derivative of this function is:

$$\partial_x \ddot{G} = \frac{1}{8\pi k^2} \int_0^\infty \begin{pmatrix} 0 & 0 & r^p k_x^3 e^{2ik_z z_0} \\ 0 & 0 & 0 \\ -r^p k_x^3 e^{2ik_z z_0} & 0 & 0 \end{pmatrix} dk_x \quad (8)$$

In the main text $\partial_x G_{xz} = \frac{1}{8\pi k^2} \int_0^\infty r^p k_x^3 e^{2ik_z z_0} dk_x$.

The imaginary part of the Green's function derivative is plotted here (*Figure S. 1*). It is clearly seen, that for the range of interest this value is always positive when integration limits include all possible modes contributions, thus the sign of the optical force is dependent on the reflection coefficient. However, if we analyze only non-evanescent modes contribution, we see a significantly weaker impact of the substrate (10 times) and a pronounced dependence on the reflection from the substrate $\text{Im}[r^p(\theta)]$. Moreover, in this case $\text{Im} \partial_x G_{xz}$ can be either negative or positive, thus optical pulling is achieved for both positive $\text{Im}[r^p(\theta)]$ and $\text{Im} \partial_x G_{xz}$ or both negative $\text{Im}[r^p(\theta)]$ and $\text{Im} \partial_x G_{xz}$. It is interesting for analysis, nevertheless, $\text{Im} \partial_x G_{xz}$ should be treated positive even for the strong surface or volumetric modes of the substrate.

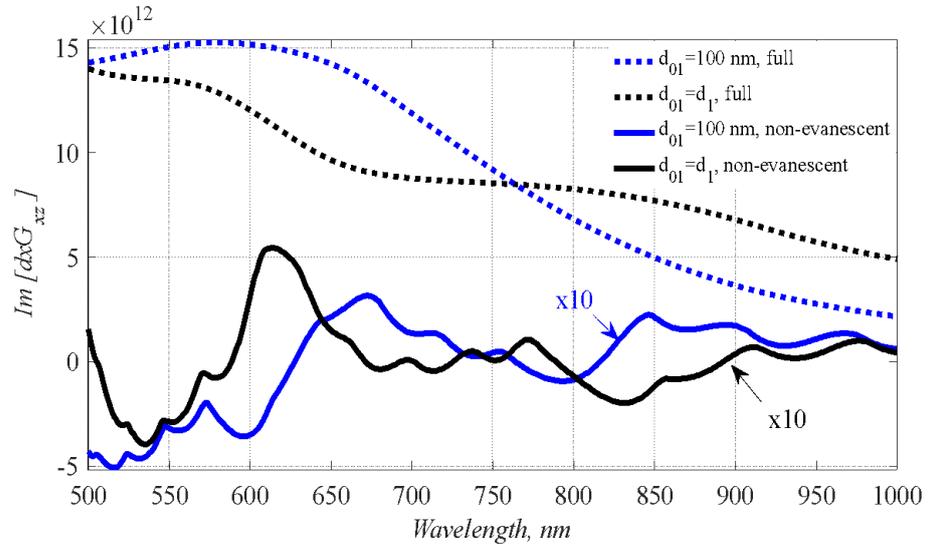

Figure S. 1 – Spectral dependence of Green's function derivative for single particle above the PC. Blue color corresponds to the specific topmost layer $d_{01} = 100$ nm, black color corresponds

to $d_{01} = d_1$. Solid lines denote integration $k_x \in [0, k]$, for visibility magnified 10 times, dashed lines denote integration over all modes $k_x \in [0, \infty)$.

## 2. Integration in the optical force

It was stated in Section III of the main text, that the BSW contribution to the optical force in the range 900-1000 nm should be distinguished precisely, because BSW lies in the narrow region between two volumetric modes ensembles. It is still possible to find contributions of the different modes in narrow spectral ranges. Thus, we show in Figure S. 2(a) that in region $k_x \in [0, A]$ exist both the free-space propagating modes and volumetric modes of the PC. For $k_x > B$ another volumetric mode's ensemble occurs. It is seen that one peak of volumetric mode takes place inside A-B region above 980 nm, however it can still be separated by changing limits of integration in this small region. By this simplified division we distinguish different modes contributions in Figure S. 2(b). The main contribution to the force as in the case demonstrated in the main paper, is given by BSW, in other words by $(A, B]$ integration region.

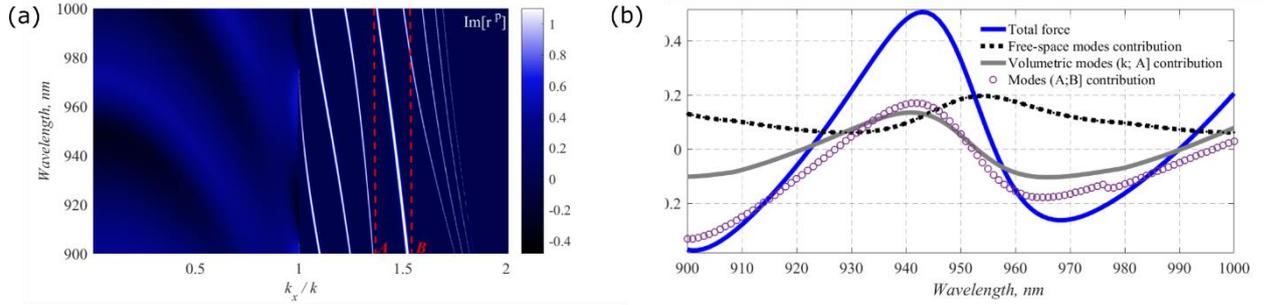

Figure S. 2 – (a) The imaginary part of the reflection coefficient. Red lines $A: k_x = 1.365k$ and $B: k_x = 1.545k$ limit BSW. (b) Optical force dependence on the incident wavelength. Black dash-dotted line depicts free-space modes contribution, grey line depicts first volumetric modes ensemble contribution $k_x \in (k, A]$ i.e., volumetric modes, violet circles shows contribution $k_x \in (A, B]$, BSW mode contribution.

## 3. Optical force near PC with non-specific topmost layer

For the topmost layer $d_{01} = d_1 = 260\,\text{nm}$, i.e., for the structure, we used to plot simplified band diagram of PC, several surface waves exist (Figure S. 3(a)). All of them contribute to the optical force (Figure S. 3(b)), however, no force enhancement is observed, because for this specific structure the width of the modes decreases. Here, the sufficient optical pulling force can be achieved for several regions. We suppose that it is possible to find parameters for the constructive interference of several BSWs and for an additional enhancement of optical forces.

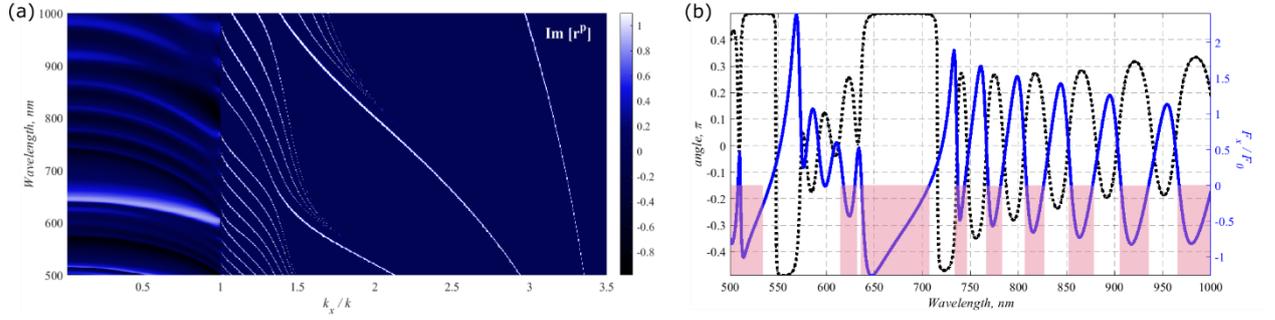

Figure S. 3 – (a) The imaginary part of reflection coefficient and (b) Optical force and dipole moment components phase difference. In the reflection coefficient colormap the three bright peaks corresponding to the BSWs excitation are clearly seen. Negative optical force corresponds to the red regions.

### 4. BSW-assisted sorting

Figure S. 4(a) shows the optical force for core-shell particles from the last section of the main paper with $R_{core} = 35, 36, 37$ nm. In Figure S. 4(b) core radius is varied $R_{core} = 35, 36.5, 37.5$ nm. The solid black line in both figures is shown as a reference and corresponds to the $R_{core} = 35$ nm, where the particle resonance occurs before main optical force extrema. For the core radius smaller than 35 nm particle's behavior is similar to the non-resonant case. As demonstrated in the main paper, changing the sign of the optical force takes place at 725 nm ($R_{core} = 35$ nm optical pulling, $R_{core} = 36$ nm optical pushing), at 750 nm ($R_{core} = 35, 36$ nm optical pulling, $R_{core} = 36.5$ nm optical pushing), at around 800 nm ($R_{core} = 35, 36, 36.5$ nm optical pulling, $R_{core} = 37$ nm optical pushing) and at around 850 and 900 nm ($R_{core} = 35, 36, 36.5, 37$ nm optical pulling, $R_{core} = 37.5$ nm optical pushing).

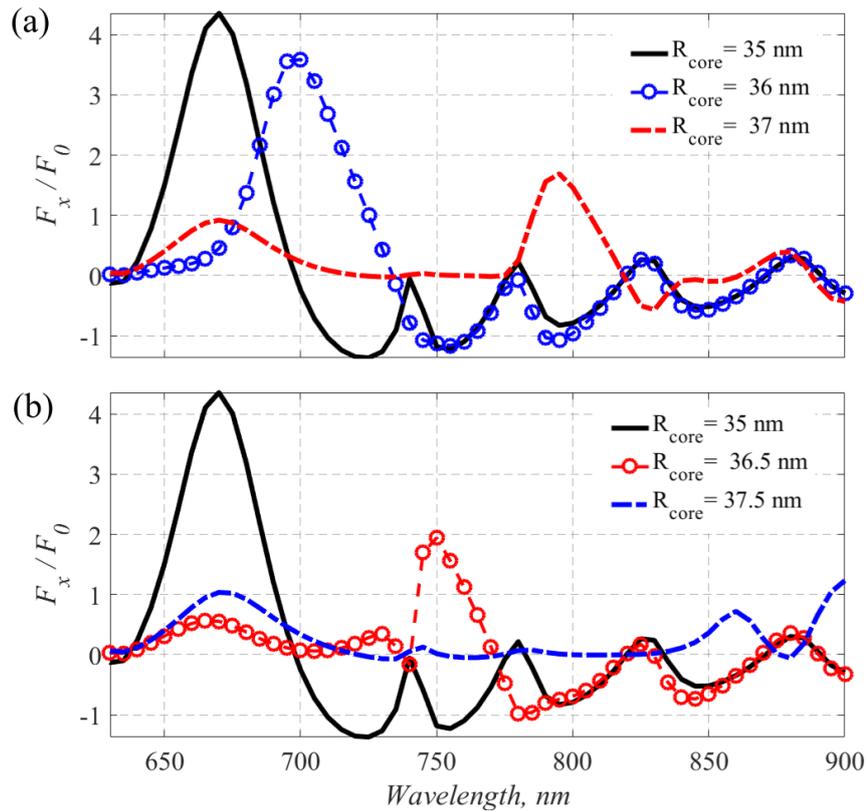

Figure S. 4 – Optical force acting on a resonant nanoparticle near the PC. (a) The blue circled line corresponds to the $R_{core} = 36$ nm, the red dashed line corresponds to the $R_{core} = 37$ nm. (b) The red circled line corresponds to the $R_{core} = 36.5$ nm, the blue dashed line corresponds to the $R_{core} = 37.5$ nm. The black solid lines show optical force for $R_{core} = 35$ nm for both figures. Insets show real parts of polarizabilities for corresponding particles without substrate contribution.